\documentclass[aps,pra,twocolumn,showpacs]{revtex4}

\usepackage{graphicx}
\usepackage{amsmath,amssymb,amsfonts}
\usepackage{natbib}

\begin{document}

\begin{abstract}
We investigate  high-order
harmonic spectra from aligned diatomic molecules in  intense driving fields whose components have orthogonal polarizations. We focus on how the driving-field ellipticity influences structural interference patterns in a macroscopic medium. In a previous publication [Phys. Rev. A \textbf{88}, 023404 (2013)] we have shown that the non-vanishing ellipticity introduces an effective dynamic shift in the angle for which the two-center interference maxima and minima occur, with regard to the existing condition for linearly polarized fields. In this work we show through simulation that it is still possible to observe this shift in harmonic spectra that have undergone macroscopic propagation, and discuss the parameter range for doing so. These features are investigated for $\mathrm{H}_2$ in a bichromatic field composed of two orthogonally polarized waves. The shift is visible both in the near- and in the far-field regime, so that, in  principle, it can be observed in experiments.
\end{abstract}

\def\andname{\hspace*{-0.5em}}
\title{Extracting an electron's angle of return from shifted interference patterns in macroscopic high-harmonic spectra of diatomic molecules}
\author{ T. Das, B. B. Augstein and C. Figueira de Morisson Faria \textit{\\Department of Physics and Astronomy, University College London,\\ Gower Street, London WC1E, 6BT, UK} \\}

\author{\ L. E. Chipperfield,  D. J. Hoffmann and J. P. Marangos \textit{\\Quantum Optics and Laser Science Group, Department of Physics, Blackett Laboratory, Imperial College London, London SW7 2BW, UK}}
\maketitle

\section{Introduction}

Recently, driving fields with nonvanishing ellipticity have attracted a great deal of attention  as potential attosecond imaging tools \cite{Kitzler_2005,Shafir_2009,Niikura_2010,Niikura_2011,Yun_2015}. In particular, with regard to molecules, these types of fields provide access to a whole parameter range which would not be accessible otherwise. For instance, they may be used to probe degenerate orbitals, molecules that are difficult to align, and  also allow the reconstruction of molecular orbitals from a single-shot measurement.  They can also be employed in molecular imaging of randomly oriented molecules \cite{Niikura_2010,Niikura_2011}.

This is made possible due to the physical mechanism behind high-order harmonic generation (HHG), namely the laser-induced recombination of an electron with its parent ion \cite{Corkum_1993,Lewenstein_1994}. By using driving fields composed of orthogonally polarized waves, one can in principle control the angle with which the electron leaves, or subsequently returns to the ion \cite{Kitzler_2005,Shafir_2009,Hoffmann_2014}. This control has been realized experimentally by employing fields with frequencies $\omega$ and $2 \omega$ and changing the relative phase between both waves \cite{Shafir_2009,Hutchison_2014}.

A particularly interesting example is  related to HHG in aligned diatomic molecules. Even if HHG is modeled within a very simple framework, namely the strong-field approximation (SFA) and the single-active electron (SAE), single-active orbital approximation, driving fields with non-vanishing ellipticity lead to interesting features. One of such features is the blurring and the splitting that occur in the two-center interference minima, which has been reported in \cite{Odzak_2010,Das_2013}. For linearly polarized fields, these minima are purely structural and have been related to electron recollision at different centers in the molecule (for the first discussion of this minimum see \cite{Lein_2002,Lein_2002_2} and for reviews see, e.g., \cite{Lein_2007} and our recent publication \cite{Brad_2012}). If, in contrast, the field is composed of orthogonally polarized waves, the interference condition changes. Indeed, in previous work, we have derived a modified two-center interference condition which incorporates the electron's angle of return as a dynamic shift \cite{Das_2013}. This shift is critically influenced by the orbit along which the electron returns, and by the harmonic energy. Hence, a coherent superposition will lead to the above-mentioned blurring and, in extreme cases, to the splitting.

A legitimate question is, however, whether this shift can be observed under more realistic conditions, such as experiments. In \cite{Das_2013} we have anticipated a series of difficulties that may arise. First, even at the single-molecule response level, there are coherent superpositions of orbits, which will be detrimental for the identification of individual shifts.  Second, the core dynamics may mask these shifts further and has not been incorporated in this model  \cite{Smirnova_2009,Smirnova_2_2009}. Third, these shifts depend on the driving-field intensity, which will vary across the beam profile. Fourth, there is no evidence that these shifts survive effects that occur during HHG propagation in a macroscopic medium.

On the other hand, propagation may facilitate the observation of these shifts, since the long and the short orbits in the dominant pair phase match differently \cite{Antoine_1996PRL,Bellini_1998}. Thus, by carefully choosing the propagation conditions one may be able to select the contributions, and the shifts, associated with individual orbits. This type of selection has been successfully used in the past to optimize attosecond-pulse production (for a seminal paper and a comprehensive review on this topic see, e.g. \cite{Antoine_1996PRL}, and \cite{Gaarde_2008}). In fact, it has even been shown that to obtain clear enough interference patterns between both orbits requires a very delicate tuning on the propagation conditions \cite{Zair_2008,Auguste_2009}. Further trajectory selection may be achieved by creating temporal or spatial gates, by playing around with the pulse shape, polarization  and/or macroscopic conditions. Examples of different trajectory-selection mechanisms are provided in the reviews \cite{Gaarde_2008,Chipperfield_2010}.

In this paper, we perform a detailed analysis of how these shifts manifest themselves in a macroscopic medium, for individual orbits. We verify that the shifts behave in different ways for the long and short orbits, and are relatively stable across the beam. This implies that they are not washed out by propagation effects, but, in principle, can be seen in a more realistic, macroscopic scenario. In our investigations we consider $\mathrm{H}_2$ to avoid multielectron effects associated with the core dynamics. We consider both linearly polarized fields and bichromatic orthogonally polarized fields of frequencies $\omega$ and $2\omega$.
For the single-molecule response, we employ the strong-field approximation (SFA) and the steepest descent method.

This paper is organized as follows. In Sec.~\ref{backgd}, we bring the necessary background to facilitate the subsequent discussion. The results are presented in Sec.~\ref{results}, from the overall behavior of the effective shifts with the driving-field intensity to an analysis of how they behave macroscopically. Finally, the main conclusions to be drawn from this paper are stated in Sec.~\ref{conclusions}.
\section{Background}
\label{backgd}
\subsection{Single-molecule response and orbit-dependent shift}
\label{singleatom}
 The SFA transition amplitude for HHG \cite{Lewenstein_1994} reads as
\begin{align}  \label{Tamp}
M(\Omega)&=-i \int_{-\infty}^{\infty}\hspace*{-0.45cm} dt \int_{-\infty}^{t}\hspace*{-0.45cm} dt^{\prime}\int \hspace*{-0.1cm}d^3
\mathbf{p} d^*_{\mathrm{rec}}(\mathbf{p}+\mathbf{A}(t)) d_{\mathrm{ion}}(\mathbf{p}+\mathbf{A}%
(t^{\prime}))  \notag \\
&\times e^{i S(t,t^{\prime },\Omega,\mathbf{p})} +c.c,
\end{align}
where $d_{\mathrm{ion}}(\mathbf{p}) =\langle \mathbf{p}|H_I(t^{\prime })|\Psi_0 \rangle$ and $d_{\mathrm{rec}}(\mathbf{p})=\langle \mathbf{p}|\mathbf{r}| \Psi_0\rangle$
are the ionization and recombination dipole matrix elements, respectively, and \begin{equation}  \label{action}
S(t,t^{\prime },\Omega,\mathbf{p})= -\frac{1}{2} \int^t_{t^{\prime }}[%
\mathbf{p}+\mathbf{A}(\tau)]^2 d\tau - I_p(t-t^{\prime}) + \Omega t
\end{equation} denotes the semiclassical action. Eq.~(\ref{action}) describes a process in which an electron propagates in the continuum from the ionization time $t^{\prime }$ to the recombination time $t$ with field-dressed momentum $\mathbf{p%
}+\mathbf{A}(\tau)$ , $t^{\prime}\leq \tau \leq t$. The ionization potential, the vector potential and the harmonic frequency are denoted by $I_p$, $\mathbf{A}$ and $\Omega$, respectively. Our computations have been performed in the length gauge, so that the interaction Hamiltonian is given by $H_I(t^{\prime})= \mathbf{r}\cdot \mathbf{E}%
(t^{\prime})$.

 For simplicity, we assume that only the highest occupied molecular orbital (HOMO) contributes to the dynamics. We also neglect the motion of the nuclei and represent the HOMO by a linear combination of atomic orbitals (LCAO). In this case, the HOMO wavefunction $\Psi _{0}(\mathbf{r})=\langle \mathbf{r}|\Psi_0\rangle$ is written as
\begin{equation}
\Psi _{0}(\mathbf{r})=\hspace*{-0.2cm}\sum_{a}c_{a }\hspace*{-0.1cm}\left[ \psi _{a }\hspace*{-0.1cm}\left( \mathbf{r}%
+\frac{\mathbf{R}}{2}\right) \hspace*{-0.1cm}+\hspace*{-0.1cm}(-1)^{\ell _{a }-m_{a }+\lambda
_{a }}\psi _{a }\hspace*{-0.1cm}\left( \mathbf{r}-\frac{\mathbf{R}}{2}\right) %
\right] \hspace*{-0.1cm},  \label{HOMOwf}
\end{equation}%
where $\psi _{a }(\mathbf{r})$ are the atomic orbitals, \textbf{R} is
the internuclear distance, $\ell_{a} $ is the orbital quantum number and $m_{a}$ is the
magnetic quantum number. The indices $\lambda_{a} =m_{a}$ relate to gerade (g) and $%
\lambda_{a} =m_{a}+1$ to ungerade (u) orbital symmetry, respectively. We incorporate the molecular structure in the prefactors and keep the action unaltered. This is a reasonable assumption if the internuclear distance is not very large (see, e.g., (\cite{Chirila_2006,Faria_2007,Faria_2009}) for details).

The transition amplitude (\ref{Tamp}) is computed using the steepest
descent method, i.e., we seek $%
t^{\prime}$, $t$ and $\mathbf{p}$ so that the action (\ref{action}) is
stationary. These stationary solutions can be associated with the tunneling time $t^{\prime}$ and the rescattering time $t$ of an electron propagating in the continuum, and it constrains its intermediate momentum $\mathbf{p}$ so that it returns to its parent ion \cite{Lewenstein_1994}. We employ either the uniform approximation or the saddle-point approximation discussed in Ref.~\cite{Faria_2002}.

We will now assume that the field is made up of two linearly polarized orthogonal
laser fields. Explicitly, the time dependent electric field $\mathbf{E}(t)=-d\mathbf{%
A}(t)/dt$ and the vector potential $\mathbf{A}(t)$ may be written as
\begin{equation}
\mathbf{E}(t)=E_{\parallel}(t)\hat{\epsilon}_{\parallel}+E_{\perp}(t)\hat{%
\epsilon}_{\perp}
\end{equation}
and
\begin{equation}
\mathbf{A}(t)=A_{\parallel}(t)\hat{\epsilon}_{\parallel}+A_{\perp}(t)\hat{%
\epsilon}_{\perp},
\end{equation}
where the field components and unit vectors along the major and the minor polarization axis are indicated by the subscripts ($||$) and ($\perp$), respectively.

The structural interference patterns are mainly given by the recombination prefactor. Here, we are interested  in $d^*_{\mathrm{rec}}(\mathbf{p}(t)\cdot \mathbf{E}(t))$, which is the complex conjugate of its component along the field-polarization direction. Explicitly,
\begin{eqnarray}
d^*_{rec}(\mathbf{p}(t)\cdot \mathbf{E}(t))\hspace*{-0.1cm}&=\hspace*{-0.1cm}%
&\sum_a\hspace*{-0.1cm}c_a\hspace*{-0.1cm}\left[ e^{-i\mathbf{%
p}(t)\cdot \frac{\mathbf{R}}{2}}+(-1)^{\ell_a-m_a+\lambda_a}e^{i\mathbf{p}(t)\cdot \frac{\mathbf{R}}{2}}\right]
\notag \\
&& \times (-i) \sum_b \partial_{p_{b}(t)}\psi^*_a(\mathbf{p}%
(t))E_{b}(t),  \label{drec2}
\end{eqnarray}
where $\mathbf{p}(t)=\mathbf{p} + \mathbf{A}(t)$, \begin{equation}
\psi_{a}(\mathbf{p}(t))=\frac{1}{(2\pi)^{3/2}}\int d^3r \psi_{a}(%
\mathbf{r})\exp[-i \mathbf{r} \cdot \mathbf{p}(t)],
\end{equation}
and b= $||$, $\perp $.

After lengthy, but straightforward computations, one may show that interference minima will occur for harmonic order
\begin{equation}
\Omega =\frac{2[n\pi -\alpha ]^{2}}{R^{2}\cos ^{2}(\theta _{L}-\zeta
(t,t^{\prime }))}+I_{p} , \label{interfmodified}
\end{equation}
where $\theta_L$ is the alignment angle of the molecule with regard to the main polarization axis, and
\begin{equation}
\zeta (t,t^{\prime })=\arctan \left[ \frac{p_{\perp }+A_{\perp }(t)%
}{p_{||}+A_{||}(t)}\right] \label{shift},
\end{equation} 
is an effective shift caused by the non-vanishing field ellipticity. In Eq. (\ref{shift}), 
\begin{equation}
p_{b}=\frac{-1}{t-t^{\prime }}\int_{t^{\prime }}^{t}A_{b}(\tau )d\tau ,  \label{pstatpar}
\end{equation}%
with $b=\parallel$ and $b=\perp$, denote the intermediate momentum of the electron along the major and minor polarization axis, respectively. 

 In other words, the angle of return has effectively been incorporated in the two-center interference condition.  The shift
$\alpha =\arctan ({-i\mathcal{A}_{+}/\mathcal{A}_{-}})$ is structural, and related to $s-p$ mixing, with
\begin{equation}
\mathcal{A}_{\pm }=\sum_a c_a\left[ (-1)^{\ell _a+m_a+\lambda _a}\pm 1\right] \eta (\mathbf{p}+\mathbf{A}(t),t)
\end{equation}%
and
\begin{equation}
\eta (\mathbf{p},t)=-i\left[ \partial _{p_{\parallel }}\psi _a^{\ast }(\mathbf{p})E_{\parallel }(t)+\partial _{p_{\perp }}\psi _a^{\ast }(\mathbf{p})E_{\perp }(t)\right].
\end{equation}
For details on this derivation we refer to our previous publication \cite{Das_2013}.
For linearly polarized fields, $\zeta (t,t^{\prime })=0$ and Eq.~(\ref{interfmodified}) reduces to the purely structural condition derived in \cite{Odzak_2009}. %

\subsection{Propagation model}
To model experimental data, we must consider the macroscopic harmonic response generated by an intense laser pulse focused into a gaseous medium, typically supplied by a gas jet directed perpendicularly across the beam. We must then numerically integrate Maxwell's wave equations with source terms distributed across this extended medium. After applying the slowly evolving wave approximation (SEWA) $\partial^{2}E/\partial z^2$ = 0, Maxwell's equations can be expressed in frequency space as \cite{Brabec_2000}
%%%%%%%%%%%%%%%%%%%%%%%%%%%%%%%%%%%%%%%%%%%%%%%%%%%%%
\begin{equation}
\frac{\partial}{\partial z}\tilde{E}(r,z,\omega) + \frac{ic}{2\omega}\nabla_{\perp}^2\tilde{E}(r,z,\omega)=- \frac{2\pi i\omega}{c}\tilde{P}(r,z,\omega),
 \label{MWE_frequency}
\end{equation}
where $\tilde{E}(r,z,\omega)$ is the electric field comprised of both the driving laser field and the generated harmonic field, and $\tilde{P}(r,z,\omega)$ is the polarisation response of the medium.
We assume that the radiation generated by nonlinear interactions does not influence the strong driving field, which is a good approximation for the parameter range of interest. This means that we can separate the total electric field into two distinct components: the driving IR laser field $\tilde{E}_l (r,z,\omega)$ and the generated XUV field $\tilde{E}_h(r,z,\omega)$, allowing us to solve Eq.~(\ref{MWE_frequency}) separately for each component using the appropriate approximations for each case.
We consider atomic gas densities between $ 10^{16}$ and $ 10^{17}$  atoms/cm$^3$ and interaction lengths of approximately a few mm. Under these conditions we can ignore all linear dispersion and absorption effects for the driving field and consider only the polarization response $\tilde{P}_{ion}(r,z,\omega)$ due to the oscillation of the free electrons created through ionization of the gas medium.

For the XUV component of the propagating field we can ignore free electron effects as these frequencies quickly exceed the plasma frequency. However, we must include the nonlinear dipole response of the molecular gas $\tilde{P}(r,z,\omega)$ and XUV absorption by the neutral atoms via the frequency dependent absorption coefficients $\alpha_{abs}$ of Ref.~\cite{Henke_1993}. In the present calculations linear dispersion due to neutral atoms has not been included. This simplification is justified for $\mathrm{H}_2$, as this molecule exhibits no resonances in the continuum for the frequency range of interest.

The contribution from the nonlinear dipole response is advanced according to
\begin{equation}
\frac{\partial}{\partial z}\tilde{E}_h(r,z,\omega)=- \frac{2\pi i\omega}{c}\tilde{P}(r,z,\omega),
 \label{nonlinear_dipole_polarisation}
\end{equation}
where
\begin{equation}
\tilde{P}(r,z,\omega)=n_a(z)M_{\tilde{\epsilon}}(r,z,\omega).
 \label{nonlinear_dipole_polarisation2}
\end{equation}
In Eq.~(\ref{nonlinear_dipole_polarisation2})$, M_{\tilde{\epsilon}}(r,z,\omega)$ is the frequency spectrum of the dipole acceleration along $\tilde{\epsilon}$, calculated using the single atom response model of Sec.~\ref{singleatom}, and $n_a(z)$ is the gas density, which we have defined using the Gaussian profile
\begin{equation}
n_a(z)= N_ae^{-4\ln 2(\frac{z-z_j}{z_\omega})^2}.
\label{Gas_Density}
\end{equation}
In the above-stated equation, $N_a $ is the peak gas density, $z_j $ marks the $z$ position of the centre of the gas jet, and $z_{\omega}$ is the full width at half maximum (FWHM) width of the jet. 

The far-field response, which is actually observed in experiments, may be obtained analytically in the frequency domain via a Huygen's integral.
This gives
\begin{align}
\tilde{E}_h(r,z+l z,\omega) =& \frac{i\omega}{cl}\int_{\infty}^{0} r^{\prime}\tilde{E}_h(r,z,\omega) e^{-\frac{i \omega}{2cl}(r^2+r^{\prime 2})} \notag \\
&J_0(\frac{rr^{\prime}\omega}{cl})dr^{\prime},
 \label{XUV absorption2}
\end{align}
where $l$ is the distance of propagation to the far field and $J_0 $ is there zeroth-order Bessel function. For more details on the propagation model we refer to our previous publication \cite{Hoffmann_2014}.

The measured XUV field is the coherent sum  of all the harmonics generated from all the atoms in the nonlinear medium. The far-field spectrum is composed of those components of the single atom dipole emission, from specific regions of the focal volume, for which there exists a wave vector along which the components add coherently \cite{Balcou_1997,Gaarde_2008}.  The degree of phase matching is usually dominated by the spatial variation of the intensity and carrier-envelope phase (CEP) of the laser field through the interaction region.  Therefore, changing the gas jet position relative to the laser focus or changing the gas jet density profile can drastically change the phase-matching conditions.  Most notably, it can be used to select either short or long plateau contributions \cite{Salieres_1995,Balcou_1997} or the half-cycle cut-off emissions \cite{Chipperfield_2010}.

\section{Results}
\label{results}
In the results that follow, we consider $\mathrm{H}_2$ in a field composed of the orthogonally polarized waves
\begin{equation}
E_{\parallel}=E_{\omega}f(t)\cos(\omega t) \label{pulsepar}
\end{equation}
and
\begin{equation}
E_{\perp}=E_{2\omega}f(t)\cos(2\omega t+\phi)\label{pulseperp},
\end{equation}
where the pulse shape $f(t)$ is a Gaussian function.
Unless otherwise stated, the fundamental pulse (\ref{pulsepar}) has wavelength $\lambda=800 $nm, peak intensity  $I_{\omega}=2.5 \times 10^{14}\mathrm{W/cm}^2$, and a full width at half maximum of 30 fs (approximately 10 cycles), which are within the experimentally relevant parameter range.

For $\mathrm{H}_2$, the HOMO is a $1\sigma_g$ orbital composed of $s$ orbitals only. Hence, the interference condition reduces to
 \begin{equation}
 \Omega=I_p+\frac{2n^2\pi^2}{R^2\cos^2\theta_L}
 \label{simplifiedinterf}
 \end{equation}
 for linearly polarized fields. For nonvanishing ellipticity, $\theta_L$ must be replaced by $\theta_L-\zeta(t,t^{\prime})$ in the above-stated equation. The orbitals employed in this work have been computed using GAMESS-UK \cite{GAMESS_2005}.
\subsection{Intensity dependence}
\label{intensity}
We will first get an insight on how the phase shift $\zeta(t,t^{\prime})$ depends on the driving-field intensity. Since this parameter varies
strongly across the laser beam profile, a simplified preliminary investigation is useful in order to understand its overall behavior.
To facilitate the interpretation,
we will approximate the pulses (\ref{pulsepar}) and (\ref{pulseperp}) by monochromatic waves.

The shifts depend strongly on the ionization and recombination times, which vary across the spectra. Specifically, there is a dominant pair of orbits, the well-known ``long orbit" and ``short orbit". For the long orbit, the electron leaves slightly after the field peak and returns after a field crossing,
while for the short orbit it ionizes at a later time and returns before the crossing. The lower the harmonic is in the plateau, the further apart
these times are. In the cutoff region, these orbits coalesce. 

\begin{figure}[tbp]
	\begin{center}
		\hspace*{-0.2cm} \includegraphics[scale=0.32]{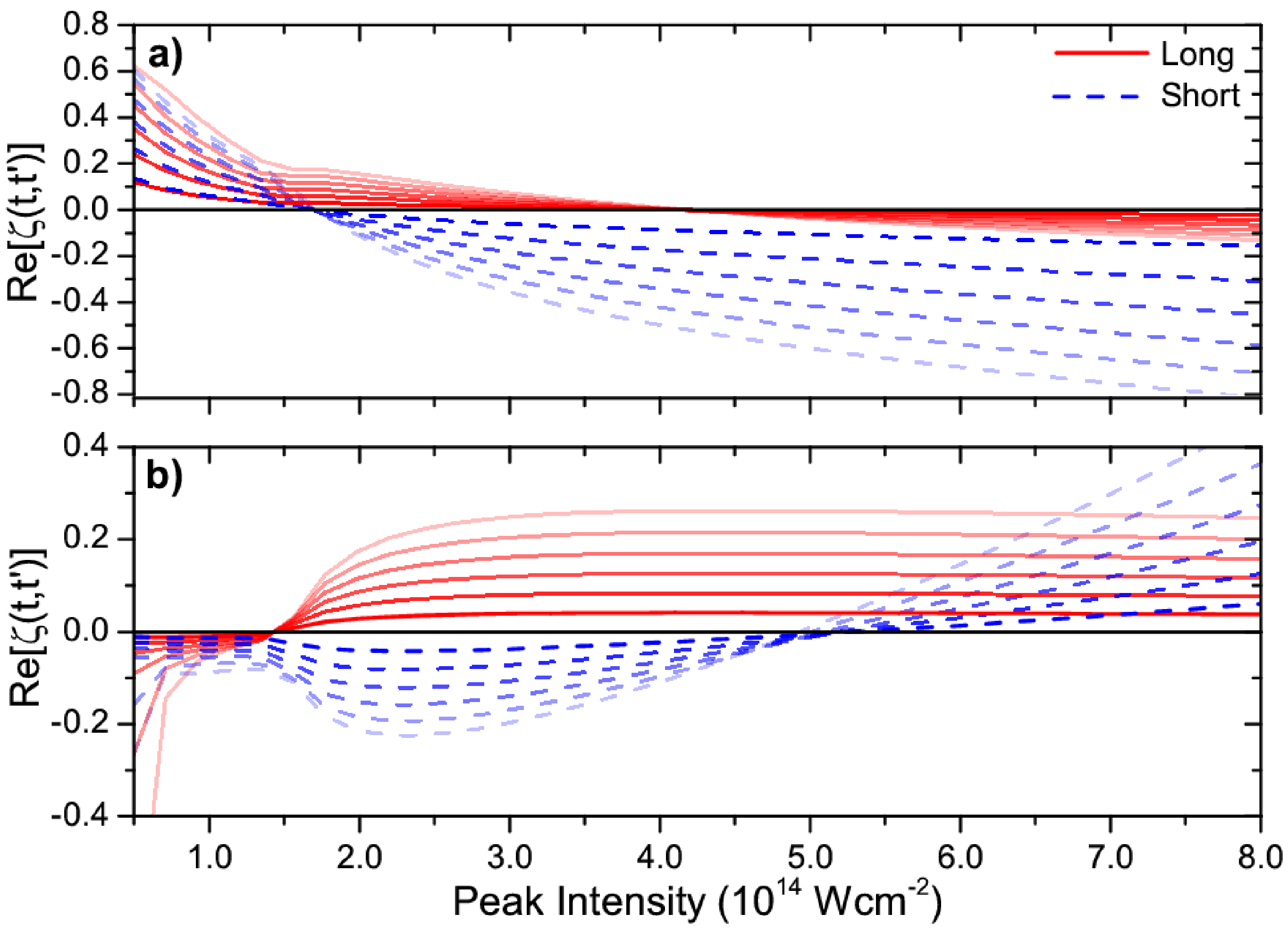}
	\end{center}
	\caption{ \label{fig:intensity1}(Color online) Real parts of the effective shifts $\protect\zeta (t,t^{\prime })$ as functions of the driving-field intensity $I_{\omega}$ of the fundamental for a harmonic of frequency $\Omega=31\omega$, using $\mathrm{H}_2$ ($I_p$= 0.59 a.u.) in two-color laser fields of increasing ellipticity.   The fundamental and the second harmonic have been approximated by monochromatic waves where $I_{2\omega}=\xi I_{\omega}$. The ellipticities have been increased from $\xi=0$ to $\xi=0.3$ in increments of $\Delta \xi=0.05$. A lighter color indicates a higher ellipticity. The dashed and solid lines refer to the short and long orbits, respectively. Panel (a) and (b) refer to relative phases of $\phi=0$ and $\phi=0.4\pi$, respectively.
	}
\end{figure}
In Fig.~\ref{fig:intensity1}, the real parts of $\zeta(t,t^{\prime})$ are plotted for a specific harmonic, as
functions of the driving-field intensity. The figure shows that the shifts depend strongly on the field ellipticity, on the phase difference $\phi$ between the two waves and on the driving-field intensity. This is expected as variations in the intensity will change the energy position occupied by a specific harmonic. This is related to the fact, that, for low enough intensities, these shifts nearly coincide for the long and short orbits. In this case, the harmonic employed is located either at or beyond the cutoff frequency. As the intensity increases, there is a splitting in the shift, indicating that the plateau region has been reached.

The behavior with regard to the field parameters depends very strongly on the orbit. If the two waves are in phase, for instance [Fig.~\ref{fig:intensity1}(a)], the distinguishing features are a large residual shift after the cutoff and a very pronounced negative shift for the short orbit in the plateau region. Physically, this means that if the electron is returning along the short orbit, for $\phi=0$ the angle of return is much larger. Hence, it may be easier to single out in a realistic scenario, especially at high intensities.

This behavior, however, changes with the relative phase. An example is provided in Fig.~\ref{fig:intensity1}(b), which exhibits a smaller residual shift in the cutoff region and comparable shifts of opposite signal if the chosen harmonic is in the high-plateau energy region. Increasing the intensity will lead to a non-trivial behavior for the shift associated with the short orbit, while the shift related to the long orbit will exhibit a monotonic behavior. Physically, this implies that the electron's angle of return will vary much more dramatically across the beam for the short orbit.
\subsection{Spatial effects across the beam profile}
\label{beam}
We will first establish how the structural minimum manifests itself in the macroscopic case, and then discuss the situation with fields of non-vanishing ellipticity. Throughout, we have chosen the propagation conditions so that the short orbit is favored. In general, this is achieved by placing the center of the gas jet after the focus. This type of configuration leads to an enhancement of on-axis harmonic emission, with low divergence. Furthermore,  we have chosen the ionization times to start shortly after the central maximum of the pulse, so that the corresponding return times will occur near the subsequent crossing.

\begin{figure}
 \includegraphics[scale=0.21]{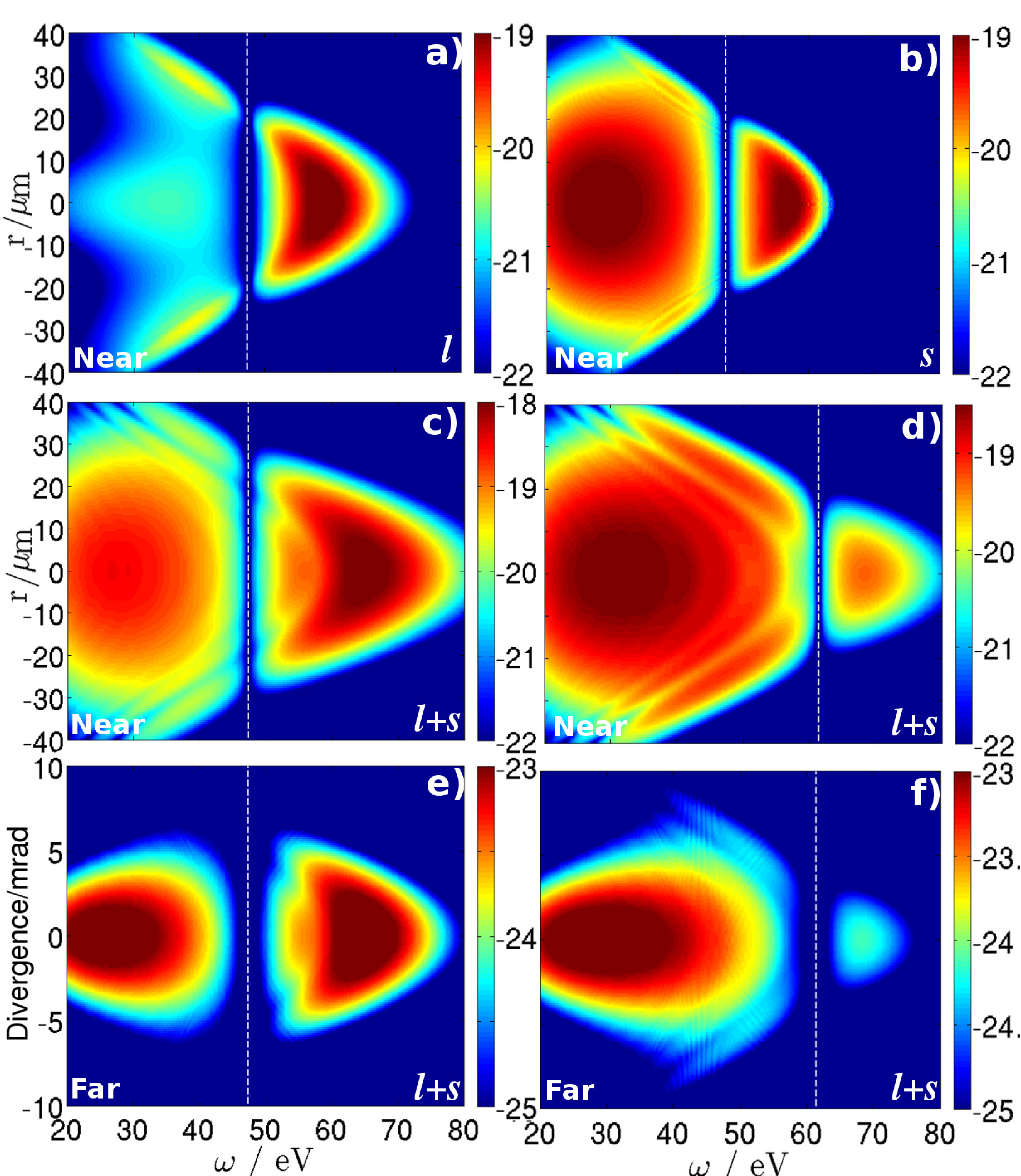}
 \caption{\label{linear} HHG macroscopic response of $H_2$ in a linearly polarized field (left and right panels, respectively) of wavelength $\lambda=800 $nm, plotted in a logarithmic scale. The beam waist is $w=30 \mathrm{\mu m}$, and the gas jet is placed at  $z_g=4$mm after the focus. The FWHM of the gas jet is 0.5mm and the FWHM of the intensity envelope is 30fs. Panels (a) and (b): Individual contributions of the long and short orbits, for a driving-field intensity $I_{\omega}=2 \times 10^{14}\mathrm{W/cm}^2$, and alignment angle $\theta=0$; note that the oscillations in Panel (b) are caused by an artifact due the inaccuracy of the standard saddle-point approximation in the cutoff region. Panels (c) and (d): spectra from the coherent superpositions of the long and short orbits, for a driving-field intensity $I_{\omega}=2.5 \times 10^{14}\mathrm{W/cm}^2$, and alignment angles $\theta=0$ [panel (c)] and  $\theta=\pi/6$ [panel (d)]; panels (e) and (f): far-field harmonic spectra, for the same intensity and alignment angles as in panels (c) and (d). The dashed line indicates the position of the structural two-center minimum. The labels $s$, $l$ and $l+s$ indicate contributions from the short orbit, long orbit, or from a coherent superposition of both, respectively.}
\end{figure}

In Fig.~\ref{linear}, we examine the structural minimum across the beam for a linearly polarized field, both near the interaction region and in the far field. This minimum is indicated by the straight vertical lines, which correspond to the interference condition Eq.~(\ref{simplifiedinterf}) and agree well with the suppression in the yield. The energy position related to this suppression remains the same regardless of whether contributions of individual orbits are taken, as shown in panels (a) and (b), or if they are combined, as shown in the remaining panels. Furthermore, it does not change with the driving-field intensity [panel (c)] or in the far-field regime [panels(e) and (f)]. Only  when the alignment angle is varied, namely for panels (d) and (f) does this minimum change. All this is consistent with the fact that the two-center minimum described by Eq.~(\ref{simplifiedinterf}) is purely structural. The fringes in panels (c) to (f) are caused by the interference of the long and short orbits.

%\begin{widetext}
 %%%%%%%%%%%%%%%%%%%%%%%%%%%%%%%%5
 \begin{figure}
 	\noindent
 	\hspace*{-0.3cm} \includegraphics[scale=0.21]{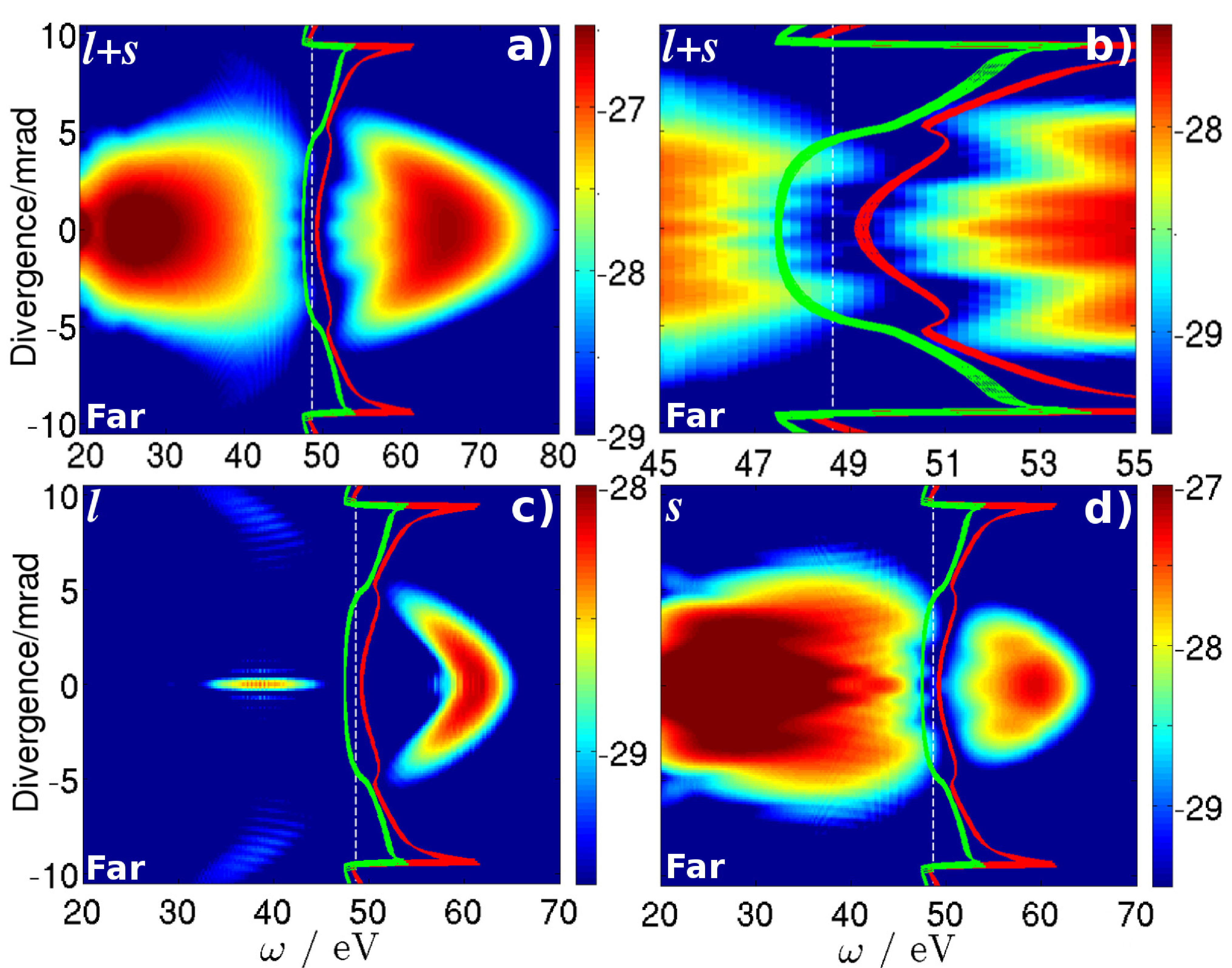}
 	\caption{\label{coherent-phi0.2} Propagated HHG spectra for $H_2$ in a Gaussian pulse, where the peak intensities of the $\omega$ and $2\omega$ waves are $ I_{\omega}=2.5 \times 10^{14}\mathrm{W/cm}^2$ and $I_{2\omega}=7.5 \times 10^{13}\mathrm{W/cm}^2$ defined at the gas jet (their intensity ratio is around 0.3) and relative phase $\phi=-0.2\pi$. The wavelength of the fundamental is $\lambda=800 nm$. The beam waist is $w=30 \mu m$ and the center of the gas jet is located at $z_g=4$ mm after the focus. The FWHM of the intensity envelope is 30fs. Panels (a) and (b) display the far-field result for a coherent superpostion of orbits denoted by the labels $l+s$, where (b) is a close-up of the region around the shifted minimum seen in (a). Panels (c) and (d) show the individual contributions to far field spectra for the long and short orbit respectively. The dashed lines and the solid lines indicate the position of the unshifted and shifted two-center minima, respectively. The red and the green solid lines give the positions of the shifted minima for the short and long orbits, respectively.  The curves have been computed spread equally across the interaction region such that $z_g-\Delta z\leq z\leq z_g+\Delta z$, with $\Delta z=0.5 mm$. All spectra have been plotted in arbitrary units and in a logarithmic scale.
 	}
 \end{figure}
 %%%%%%%%%%%%%%%%%%%%%%%%%%%%%%
 
 In the upper panels of Fig.~\ref{coherent-phi0.2} we present the macroscopic HHG spectrum computed for orthogonally polarized fields considering a coherent superposition of the long and short orbits, and a phase difference of $\phi=-0.2\pi$ between the $\omega$ and $2\omega$ waves. This figure illustrates a particular case for which the shift associated with the short orbit, given by the red lines, is visible in a realistic scenario, and survives in the far field. The close-up near the interference minimum [Fig.~\ref{coherent-phi0.2}(b)] shows that it follows the generalized interference condition related to the short orbit very closely. Only at the beam edges there are small discrepancies from this condition, and the suppression approaches the shifted minimum associated to the long orbit, which is given by the green line.
  
The reason behind this clear picture can be seen in panels (c) and (d) of Fig.~\ref{coherent-phi0.2}, in which we present the individual contributions from the long and short orbits. They show that the contributions from the long orbit are strongly suppressed in the plateau region, and that those from the short orbit are much more significant. This stems from the fact that the relative phase $\phi=-0.2\pi$ selects the short orbit at the single-molecule level via polarization gating \cite{Hoffmann_2014}. This selection is then reinforced by the appropriate propagation conditions. Also for that reason there are no visible fringes when the coherent superposition of both orbits is considered (Fig.~\ref{coherent-phi0.2}). 

\begin{figure}
	\hspace*{-0.3cm} \includegraphics[scale=0.21]{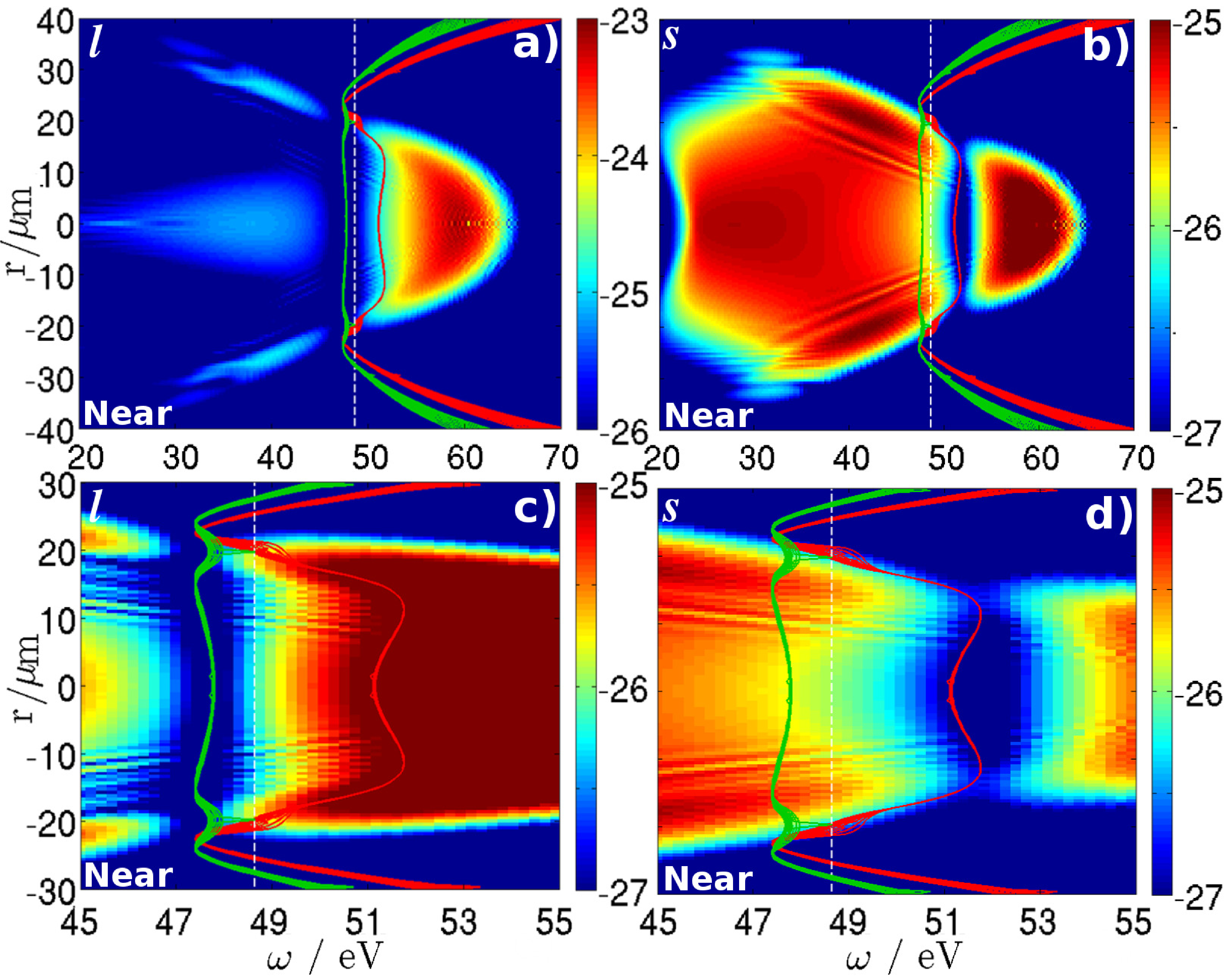}
	\caption{\label{elliptical-phi0}Individual contributions of the long and short orbits (left and right panels, respectively) for the HHG macroscopic response of $H_2$ in a Gaussian pulse composed of parallel and perpendicular waves given by Eq.~(\ref{pulsepar}) and (\ref{pulseperp}), respectively. The peak intensities of the $\omega$ and $2\omega$ waves are $ I_{\omega}=2.5 \times 10^{14}\mathrm{W/cm}^2$ and $I_{2\omega}=7.5 \times 10^{13}\mathrm{W/cm}^2$ defined at the gas jet (their intensity ratio is around 0.3) and relative phase $\phi=0$. The center of the gas jet is located at $z_g = 2mm$ after the focus. This phase has been found to give a very large residual shift $\zeta(t,t^{\prime})$ at the cutoff. The wavelength of the fundamental is $\lambda=800 nm$. The beam waist is $w=30 \mu m$, and the FWHM of the intensity envelope is 30fs. Panels (a) and (b): HHG yield from the individual orbits; panels (c) and (d): zoom in of the upper panels close to the interference minimum. The dashed lines and the solid lines in the upper and middle panels indicate the position of the unshifted and shifted two-center minima, respectively. The red and the green solid lines give the positions of the shifted minima for the short and long orbits, respectively. The curves have been computed spread equally across the interaction region such that $z_g-\Delta z\leq z\leq z_g+\Delta z$, with $\Delta z=1 mm$. All panels have been plotted in arbitrary units and in a logarithmic scale. The labels $l$ and $s$ are associated to the long and the short orbit, respectively.}
\end{figure}

\begin{figure}
	\noindent
	\hspace*{-0.3cm} \includegraphics[scale=0.21]{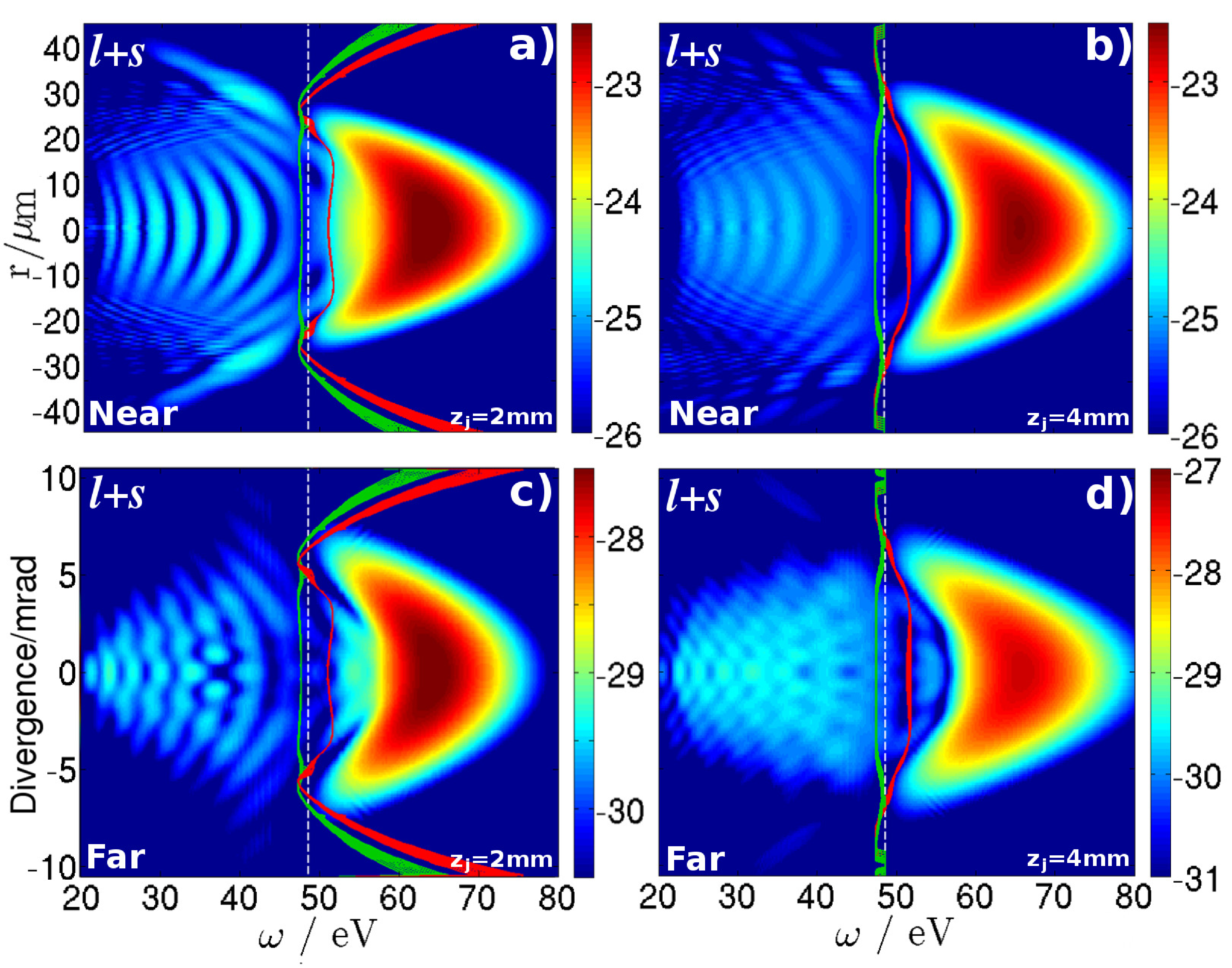}
	\caption{\label{coherent-phi0} Propagated spectra considering a coherent superposition of the long and short orbits, for the same field parameters as in Fig.~\ref{elliptical-phi0}. Most propagation conditions have also been kept as in Fig.~\ref{elliptical-phi0} except the center of the gas jet, which has been chosen to be at  $z_g = 2mm$ and $z_g = 4mm$ after the focus (left and right panels, respectively). Panels (a) and (b) display the spectra in the interaction region, while panels (c) and (d) show the far-field results. The dashed lines and the solid lines indicate the position of the unshifted and shifted two-center minima, respectively. The red and the green solid lines give the positions of the shifted minima for the short and long orbits, respectively.  The curves have been computed spread equally across the interaction region such that $z_g-\Delta z\leq z\leq z_g+\Delta z$, with $\Delta z=1 mm$. All spectra have been plotted in arbitrary units, and in a logarithmic scale.
	}
\end{figure}

 Next, we will analyze the generalized interference condition for individual orbits in more detail. With that aim in mind, we (i) choose a phase difference $\phi=0$, for which these shifts are expected to be large \cite{Das_2013}; (ii) consider the near-field regime so that diffraction effects are ruled out. These results are plotted in Fig.~\ref{elliptical-phi0}, for the short and long orbits. The figure shows that these shifts depend very strongly on the orbit. While, for the long orbit, the shift moves the minimum away from the cutoff and towards lower frequencies, for the short orbit it causes it to move towards the cutoff [see panels (a) and (b)]. Furthermore, a zoom-in of the shifts, presented in panels (c) and (d), shows that they approach each other around the cutoff and follow the modified interference condition (\ref{interfmodified}) very closely. In particular, the shift is much larger for the short orbit and, as long as the cutoff has not been reached, it varies very little within the interaction region. For the long orbit, the shift varies slightly more with regard to $z$. Beyond the cutoff, the variation is more extreme. This region, however, is not of interest to the present problem.

Unfortunately, however, these shifts cannot be seen if a coherent superposition of orbits is considered. We have plotted this superposition in Fig.~\ref{coherent-phi0}, for the near and far field regimes [upper and lower panels in the figure, respectively]. The overall behavior shows many interference fringes and no clear shifted minimum. This happens because, at the single-molecule response level, a phase $\phi=0$ will enhance the long orbit and suppress the short orbit \cite{Hoffmann_2014}. Hence, there are conflicting conditions from the the single-molecule response and propagation, which is not ideal. This is clearly seen in panels (a) and (c) of the figure, which exhibit fringes in the whole plateau region due to the interference between the short and long orbits. Close to the two-center minimum the results are inconclusive as (i) the suppression near the green curve seems much more related to the interference between the long and short orbit than to the electron's angle of return being incorporated; (ii) the shift associated to the short orbit does not manifest itself as a clear suppression. Both issues (i) and (ii) become slightly better in the far-field, as the interference between the short and long orbit is partially washed out. Nonetheless, the result is not as clear as in Fig.~\ref{coherent-phi0.2}, for which the short orbit was favored at the microscopic and macroscopic level. 

This problem may be attenuated by moving the gas jet further away from the focus, in order to favor the short orbit [see Figs.~\ref{coherent-phi0}(b) and (d)]. In this case, the interference fringes in the plateau become more blurred, so that the shift related to the short orbit can be identified.

One should bear in mind, however, that we have restricted the contributing orbits to a single half cycle of the driving field. This does not correspond to a realistic situation, as a real pulse would have many cycles. A specific problem is that, in a bichromatic $\omega-2 \omega$ field with orthogonally polarized waves, the long and short orbits starting at the subsequent half cycle gives rise to shifts $\zeta(t,t^{\prime})$ of opposite signs \cite{Das_2013}. In practice this causes a blurring in the generalized interference condition. This problem may be overcome if one employs a few-cycle pulse.

\begin{figure}
\noindent
\hspace*{-0.3cm} \includegraphics[scale=0.22]{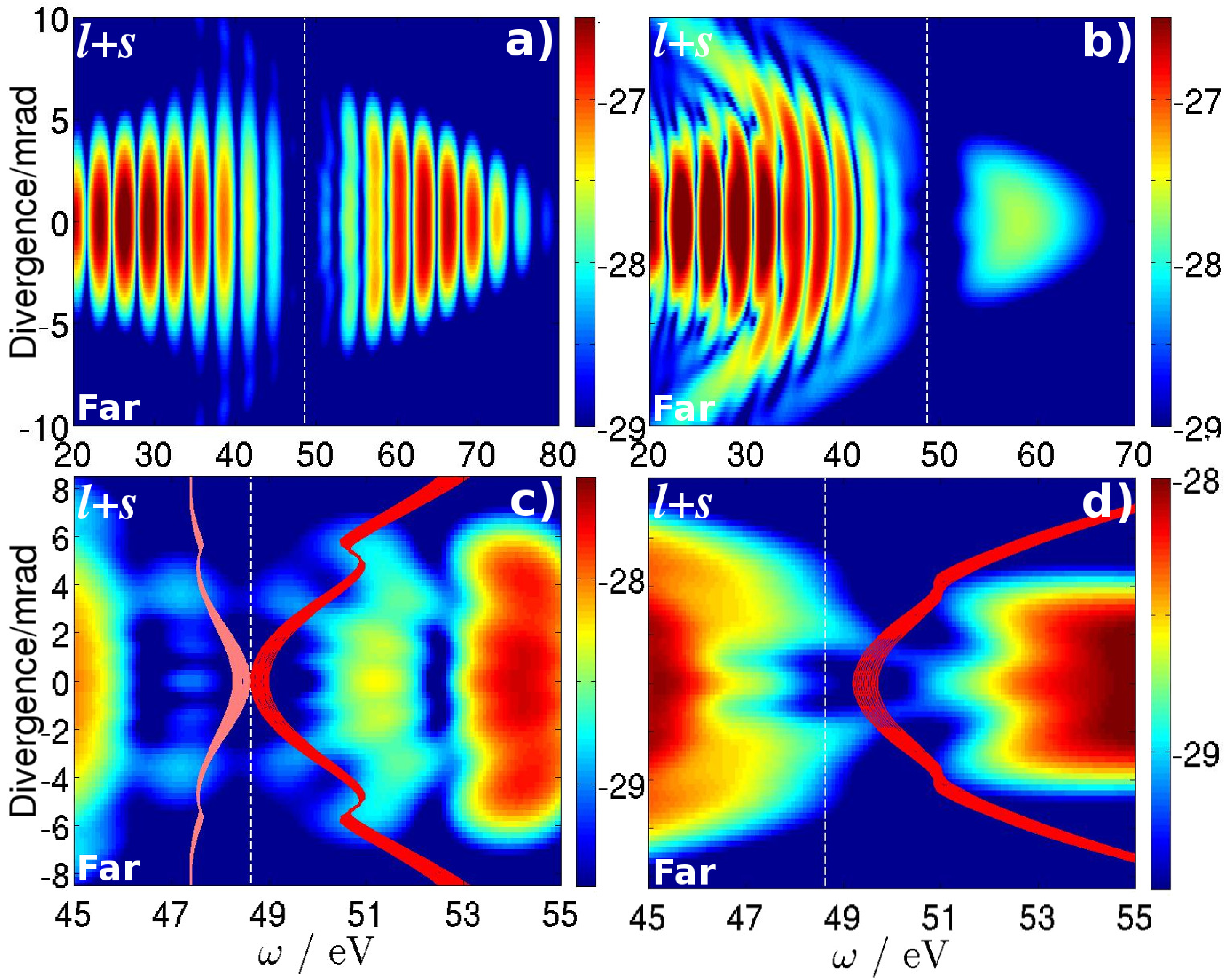}
\caption{\label{fig:fewcyclevs mono}Propagated HHG spectra considering two half cycles, for the same driving-field parameters in Fig.~\ref{coherent-phi0.2}, but different FWHM of the intensity envelope. In panels (a) and (c), this width is 30fs, while in panels (b) and (d) it is 5.5fs. In the lower panels we show only the shift related to the short orbit, which dominates throughout. These shifts are displayed as the thick lines in the figure. The shifts associated with the first and second half cycle are shown as the red curves in panels (c) and (d), and the pink curve in panel (c), respectively. }
\end{figure}

This is shown in Fig.~\ref{fig:fewcyclevs mono}, in which we present the far-field spectra calculated using the two dominant half cycles of the pulse in Fig.~\ref{coherent-phi0.2}, and of a few-cycle pulse with the same parameters, except the full width at half maximum of the intensity envelope (left and right panels, respectively). For the long pulse, the contributions from both half cycles blur the minimum associated with the short orbit [panel (c)], while for the few-cycle pulse it remains very clear.  It is also well described by the interference condition (\ref{interfmodified}) [panel (d)].
This happens because the two-center minimum is in the plateau for the first half cycle, while it is beyond the cutoff for the second half cycle of the short pulse. The different cutoff energies become clear in panel (b), which shows well-defined harmonics only for energies lower than the structural minimum. In contrast, the long pulse leads to well-defined harmonics throughout, as seen in panel (a). 

\section{Conclusions}
\label{conclusions}
In this work, we have performed theoretical studies of the macroscopic response of high-harmonic radiation from diatomic molecules in bichromatic fields composed of two orthogonally polarized driving waves. Our computations show that, for an appropriate choice of driving field and propagation conditions, the angle of return of an electron to its parent molecule manifests itself as a dynamic shift in the two-center structural minimum. This shift depends on the driving-field intensity, on the harmonic order and on the orbit along which the electron returns. For the dominant orbits, these shifts are visible both in the near- and far-field regimes, indicating that they can, in principle, be measured in experiments.

In order to see these shifts clearly, one must select an individual orbit by employing gating mechanisms at the single molecule level and by using appropriate propagation conditions. It is of particular interest to enhance the short orbit and suppress the long orbit, as it phase matches on axis and leads to more dramatic variations in the angle of return with the laser-field parameters. If conflicting microscopic and macroscopic conditions are provided, the interference between long and short orbits may lead to inconclusive results. With that aim in mind, we have selected the phase difference between both $\omega$ and $2\omega$ driving waves as $\phi=-0.2\pi$ and have placed the focus before the gas jet. These choices are known to enhance the short orbit at the single-atom \cite{Hoffmann_2014} and macroscopic \cite{Balcou_1997,Chipperfield_2010} levels, respectively.

Nonetheless, there exist two obstacles towards seeing this shift, which, however, are well under control. First, for bichromatic fields, the shifts coming from the other half cycle will have opposite sign. This flip may be avoided for few-cycle pulses, by creating a spectral and temporal gate in order to select electron orbits from a single half cycle \cite{Chipperfield_2010,Haworth_2007}. This can be performed if for one of the dominant half cycles, the shifted minimum lies beyond the cutoff energy, and within the plateau for the other half cycle, as shown in Fig.~\ref{fig:fewcyclevs mono}. 

Second, for molecules with a more complex electronic structure, the core dynamics may also play an important role, such that the single-active orbital, single-active electron approximation is not applicable \cite{Smirnova_2009}. The regime for which this happens, however, can be avoided. Indeed, recent experiments in $\mathrm{CO}_2$ have shown that interference effects stemming from the core dynamics are only relevant if the structural minimum provides a window for them to be observed. For high enough intensities, the energy position of the dynamical minimum lies outside this window and can no longer be observed \cite{Torres_2010}. This is in agreement with  the experimental findings in \cite{Kato_2011}, which support the structural instead of the dynamical interference picture, and with the computations in \cite{Rupenyan_2013}, which show that multielectron effects are important in the cutoff, but not in the mid-plateau energy region.

Finally, our results suggest that, in principle, the angle with which the electron returns to the core can be controlled, either by changing the relative phase between both waves, or the driving-field intensity. As the intensity increases, a specific harmonic moves from the cutoff region across the plateau towards the ionization threshold, and this will alter the angle of return. This behavior is particularly critical for the short orbit, which, as discussed above, is the most favorable for macroscopic observations of this shift. Furthermore, this angle can be mapped into a shift near the structural minimum, which can be modified by an appropriate parameter choice. This may be desirable in future experiments in order to investigate dynamic effects which would be obfuscated otherwise.
\section*{Acknowledgements}
 This work has been supported by UCL (Impact PhD studentship) and by the Engineering and Physical Sciences Research Council (UK) (EPSRC) (grants EP/I032517/1 and EP/J019240/1 and Doctoral Training Prize) and the ERC ASTEX project 290467. The authors acknowledge the use of the UCL Legion High-Performance Computing Facility (Legion@UCL), and associated support services, in the completion of this work. We are especially indebted to M. Uhrin for his help with the coding and debugging, and to H. Kelly for her help with adapting the propagation code to the high-performance computers at UCL.

\end{document}